\documentclass[aps,preprint,superscriptaddress]{revtex4}

\usepackage{natbib}
\usepackage{graphicx}
\usepackage{amsmath}
\usepackage{amssymb}
\usepackage{color}

\begin{document}

\title{Wavelet Methods for Studying the Onset of\\ Strong Plasma Turbulence}

\author{A. Le}
\affiliation{Los Alamos National Laboratory, Los Alamos, New Mexico 87545, USA}
\author{V. Roytershteyn}
\affiliation{Space Science Institute, Boulder, CO 80301, USA}
\author{H. Karimabadi}
\affiliation{Analytics Ventures, Inc., San Diego, CA 92121, USA}
\author{A. Stanier}
\affiliation{Los Alamos National Laboratory, Los Alamos, New Mexico 87545, USA}
\author{L. Chacon}
\affiliation{Los Alamos National Laboratory, Los Alamos, New Mexico 87545, USA}
\author{K. Schneider}
\affiliation{Institut de Math\'ematiques de Marseille (I2M), Aix-Marseille Universit\'e, 13284 Marseille, France}

\begin{abstract}
Recent simulations have demonstrated that coherent current sheets dominate the kinetic-scale energy dissipation in strong turbulence of magnetized plasma. Wavelet basis functions are a natural tool for analyzing turbulent flows containing localized coherent structures of different spatial scales. Here, wavelets are used to study the onset and subsequent transition to fully developed turbulence from a laminar state. Originally applied to neutral fluid turbulence, an iterative wavelet technique decomposes the field into coherent and incoherent contributions. In contrast to Fourier power spectra, finite time Lyapunov exponents (FTLE), and simple measures of intermittency such as non-Gaussian statistics of field increments, the wavelet technique is found to provide a quantitative measure for the onset of turbulence and to track the transition to fully developed turbulence. The wavelet method makes no assumptions about the structure of the coherent current sheets or the underlying plasma model. 
Temporal evolution of the coherent and incoherent wavelet fluctuations is found to be highly correlated (Pearson correlation coefficient of $ > 0.9$) with the magnetic field energy and plasma thermal energy, respectively. 
The onset of turbulence is identified with the rapid growth of a background of incoherent fluctuations spreading across a range of scales and a corresponding drop in the coherent components. This is suggestive of the interpretation of the coherent and incoherent wavelet fluctuations as measures of coherent structures (e.g., current sheets) and dissipation, respectively. The ratio of the incoherent to coherent fluctuations $R_{ic}$ is found to be fairly uniform across different plasma models and provides an empirical threshold for turbulence onset. 
The utility of this technique is illustrated through examples. First, it is applied to the Kelvin--Helmholtz instability from different simulation models including fully kinetic, hybrid (kinetic ion/fluid electron), and Hall MHD simulations. Second, the wavelet diagnostic is applied to the development of turbulence downstream of the bowshock in a global magnetosphere simulation. Finally, the wavelet technique is also shown to be useful as a de-noising method for particle simulations.
\end{abstract}

\maketitle


\section{Introduction}

While many studies have focused on methods to describe the properties of fully developed turbulence, much less attention has been paid to techniques to describe the onset and subsequent transition to fully developed turbulence. The aim of the present study is to address this shortcoming. 

Large-scale plasma turbulence is understood to involve formation of localized, coherent current sheets at different spatial scales.  These coherent structures appear to dominate the energy dissipation of turbulence on kinetic scales as compared to other dissipation mechanisms such as wave interactions~\cite[e.g.,][]{wan:2012,karimabadi:2013,karimabadi:2014}. While the plasma waves are naturally analyzed in terms of Fourier modes, localized structures call for decompositions that reflect their localization and their multi-scale properties.


Wavelets, which are localized functions in space and scale, offer the possibility to represent intermittent functions and localized structures exhibiting a large range of scales in an efficient way.
The so-called `mother wavelet', $\psi(x)$, which has finite energy, is the elementary building block of the wavelet transform. It is a well-localized function with fast decay at infinity and at least one vanishing moment ({\it i.e.}, zero mean) or more. It is also sufficiently smooth, which implies that its Fourier transform exhibits fast decay.
The wavelet transform introduced in~\cite{grossmann:1984} decomposes a signal ({\it e.g.}, in time) or any field ({\it e.g.}, in three-dimensional space) into both space (or time) and scale (or time scale), and possibly directions (for dimensions higher than one). 

Wavelets have been used for analyzing hydrodynamic turbulence starting in the 1990s and then extended for modeling and computing turbulent flows (see review articles \cite{farge:1992}, \cite{schneider:2010}).
Here, we provide a brief summary of application of wavelet techniques in the context of plasma turbulence. Early examples include use of wavelets in analysis of space \cite{dudok:1995,voros:2004} and laboratory \cite{vanmilligen:1995} plasmas. Wavelet filtering has been used for extracting coherent bursts in turbulent ion density plasma signals, measured by a fast reciprocating Langmuir probe in the scrape-off layer of the tokamak Tore Supra (Cadarache, France) \cite{farge:2006}.
Wavelet-based density estimation techniques have also been used to improve particle-in-cell numerical schemes \cite{nguyen:2010}, and a particle-in-wavelet scheme was developed for solving the Vlasov--Poisson equations directly in wavelet space \cite{NSSF11}.
Wavelet de-noising has been applied for tomographic reconstruction of tokamak plasma light emission in \cite{NFBBSFM12}.
Coherent Vorticity and Current sheet Simulation (CVCS), which applies wavelet filtering to the resistive non-ideal MHD equations, was proposed as a new model for turbulent MHD flows.
It allows a reduction in the number of degrees of freedom necessary to compute the flows, while capturing the nonlinear dynamics of the flow.
Recently Groselj \textit{et al.} \cite{groselj:2018} analyzed high-resolution observational data and state-of-the-art numerical simulations to study the relationship between wavelike physics and large-amplitude structures in astrophysical kinetic plasma turbulence using the continuous wavelet transform with complex valued wavelets.
A review on wavelet transforms and their applications to MHD and plasma turbulence can be found in Ref.~\cite{farge:2015}.

The aim of the present paper is to use the orthogonal wavelet decomposition of turbulent flows into coherent and incoherent contributions to define a criterion that determines the onset of plasma turbulence. To this end, high-resolution numerical simulations of unstable shear-flows triggered by the Kelvin--Helmholtz instability using different approaches---fully kinetic, hybrid (kinetic ion/fluid electron), or Hall MHD---are analyzed using orthogonal wavelets. This technique is then further tested in a more complex scenario of turbulence generation downstream of the bow shock in a global hybrid simulation of the magnetosphere. Comparison with Fourier power spectra and non-Gaussianity diagnostics is presented.

Temporal evolution of the coherent and incoherent wavelet fluctuations is found to be highly correlated (Pearson correlation coefficient of $> 0.9$) with the magnetic field energy and plasma thermal energy, respectively. This is suggestive of the interpretation of the coherent and incoherent wavelet fluctuations as measures of coherent structures (e.g., current sheets) and dissipation, respectively. Since plasma heating can be partly due to reversible processes (e.g., adiabatic), a more rigorous connection between the incoherent fluctuations and dissipation will be explored elsewhere and is beyond the scope of this work.

The outline of the paper is the following: In Section~\ref{sec:fourwave} we recall Fourier and wavelet analysis and in Section~\ref{sec:iterative} the iterative wavelet filtering is presented. The simulation set-ups are described in Section~\ref{sec:sim}.  Section~\ref{sec:transit} introduces a wavelet-based method for quantifying the transition of flows to turbulence and compares and contrasts it with three more traditional techniques for studies of turbulence, Fourier power spectra, structure function, and finite time Lypaunov exponent (FTLE). The wavelet method is then applied to a more complex flow in a global magnetosphere simulation in Section~\ref{sec:global}, and a summary is given in Section~\ref{sec:summary}. The Appendix discusses and demonstrates the use of the wavelet technique for de-noising particle simulations.

\section{Fourier and Wavelet Analysis}
\label{sec:fourwave}

In the following, we review a few concepts related to Fourier and wavelet analysis in the context of studying turbulent plasma flows. Fourier modes arise naturally in the study of weak plasma turbulence. Because the full non-linear equations of motion of a plasma (in kinetic and fluid descriptions) are analytically intractable, much analytic work has focused on the linear approximation. For homogeneous plasmas, weak fluctuations are then typically described by normal modes that vary as independent Fourier components $\propto \exp(i\omega t - ikx)$, with a dispersion relation $\omega = \omega(k)$ imposed by the linearized equations of motion. Theories of weak plasma turbulence were developed by treating the non-linear interactions of the normal modes by perturbation theory \cite{sturrock:1957,galeev:1965,kennel:1966,galtier:2000}. The complexity of turbulent flows is then ascribed to the interaction of a large number of incoherent Fourier components \cite{manneville:1995}, resulting in a cascade of energy \cite{kolmogorov:1941} between large scales (low $k$) and small scales (high $k$).

While Fourier analysis is thus suited for studying weak turbulence, it may not be well-adapted for characterizing strongly non-linear flows. Strongly turbulent fluid flows include coherent structures such as vortex tubes \cite{farge:2001}, while magnetized plasma turbulence displays the formation of current sheets \cite{politano:1995,tenbarge:2013,zhdankin:2013,karimabadi:2013}. In Fourier space, these localized coherent structures require a large number of modes for their description. As described below, wavelets yield a sparse representation of intermittent data.

\begin{figure}
\includegraphics[width = 0.8\textwidth]{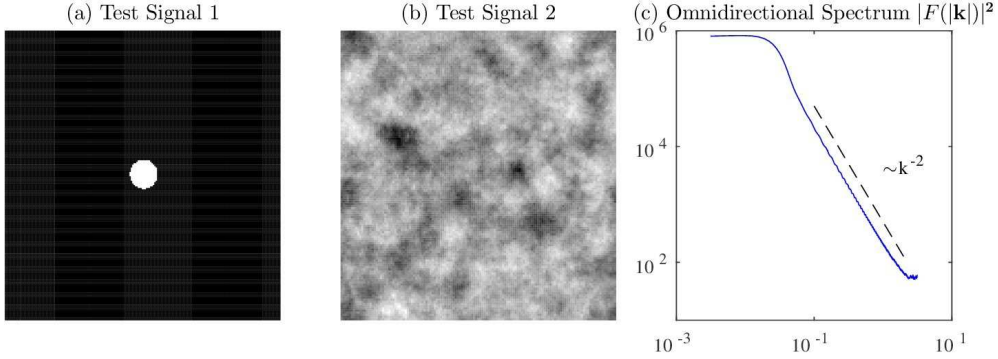}
\caption{(a-b) Two test signals have identical (c) Fourier power spectra (these spectra are integrated over angular directions and use finite bins in $k$). Fourier modes are spread over space, and the localized structure in (a) therefore consists of coherent contributions from nearly all Fourier modes. \label{fig:coherent}}
\end{figure}

To illustrate a limitation of Fourier spectral analysis, two test signals are shown in Figs.~\ref{fig:coherent}(a) and (b) that have identical Fourier power spectra, while their Fourier modes have different phases. 
In terms of Fourier wave number, each signal exhibits a power-law tail that scales as $|\tilde{F}(k)|^2\propto 1/|k|^2$. It is apparent, however, that the two signals are qualitatively different. Test Signal 1 in Fig.~\ref{fig:coherent}(a) is completely localized to a central circle, whereas Test Signal 2 in Fig.~\ref{fig:coherent}(b) is spread over space. Because Fourier modes extend over all of space, capturing a localized signal requires coherent contributions from a large number of Fourier modes. As a 1D example, the Heaviside step function, which is discontinuous and defined by
\begin{equation}
\theta(x) = \begin{cases} 0, & \mbox{if } x\le0 \\ 1, & \mbox{if } x>0 \end{cases}
\end{equation}
has Fourier components $\tilde{\theta}(k)=-i/2\pi k$ for $|k|>0$. The sharp jump in $\theta(x)$ is thus encoded in the coherent phases of its Fourier components, which display a power-law as a function of $k$, similar to the 2D example of Test Signal 1. By contrast, Test Signal 2's Fourier power spectrum also contains a $1/k^2$ tail, but there is little information contained in the coherence of the phases of the Fourier components. Indeed, Signal 2 was generated by multiplying each Fourier component of Signal 1 by a (pseudo-)random complex phase.

Wavelets, see e.g., Ref.~\cite{mallat:1989}, provide a different set of basis functions, and they are particularly adapted for capturing localized coherent structures across a range of different scales. Like Fourier modes, certain wavelet families form ortho-normal bases for decomposing functions of one or more variables. Unlike Fourier modes, however, wavelets are not solutions to any particular physical equations. Rather, they are constructed to analyze general signals with multi-scale structure. While Fourier modes capture a single wavelength but are spread out over physical space, wavelets encode localized information about both scale and position.

\begin{figure}
\includegraphics[width = 0.5\textwidth]{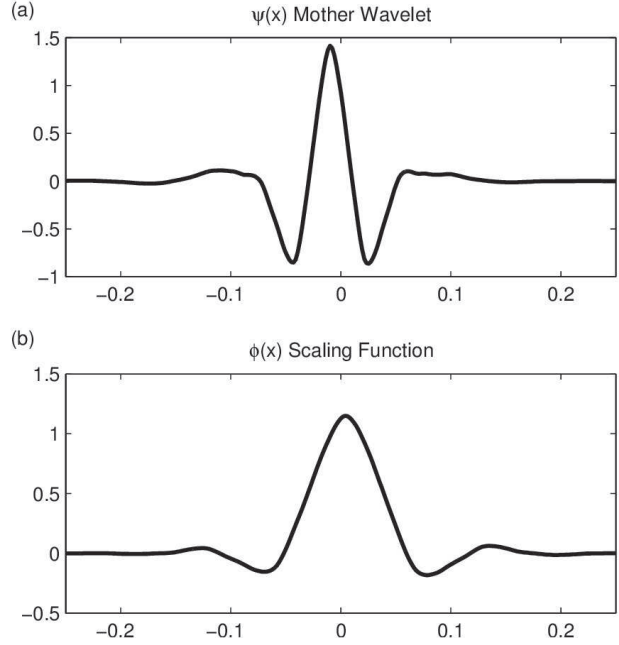}
\caption{(a) The mother wavelet of the coiflet-18 wavelet family used in our analysis. (b) The corresponding coiflet scaling function.\label{fig:coif}}
\end{figure}

Several different wavelet families have been derived. Here, we use a discrete "coiflet-18" basis \cite{daubechies:1993} that gives rise to a multi-resolution \cite{jawerth:1994} representation of 2D functions. The family of wavelets is built out of two specially chosen functions: a so-called mother wavelet $\psi(x)$ and a scaling function $\phi(x)$, each of which is plotted in Fig.~\ref{fig:coif}. One key characteristic of this wavelet function $\psi(x)$ is its compact support, i.e., it is non-zero over only a finite range. The family of 1D wavelets is given by the translations and dilations of the mother wavelet:
\begin{equation}
\psi_{l,n}(x) = \frac{1}{\sqrt{2^l}}\psi\left(\frac{x-2^ln}{2^l}\right),
\end{equation}
where the index $l$ and shift $n$ each span the integers. Built from these 1D wavelets along with similar translations and dilations of the scaling function $\phi$, an ortho-normal basis for 2D functions may be defined by 
\begin{equation}
\Psi^p_{l,m,n}(x,y) = \begin{cases}
\psi_{l,m}(x)\phi_{l,n}(y) , & \mbox{for } p=H \\
\phi_{l,m}(x)\psi_{l,n}(y) , & \mbox{for } p=V \\
\psi_{l,m}(x)\psi_{l,n}(y) , & \mbox{for } p=D \\
\end{cases}
\end{equation}
where again $l$, $m$, and $n$ span the integers; and $p$ corresponds to three directions typically referred to as $H$orizontal, $V$ertical, and $D$iagonal. In practice for our discrete simulation data, we decompose each signal over a finite number of levels $l<L$ and shifts (which depend on the size of the numerical grid and the level). Up to corrections for boundary cells, each 2D field $F(x,y)$ defined on the computational grid is de-composed in the wavelet basis as:
\begin{equation}
F(x,y) = \sum_{m,n} A_{m,n}\phi_{L,m}(x)\phi_{L,n}(y)  + \sum_{m,n,l,p}D_{p,l,m,n}\Psi^p_{l,m,n}(x,y),
\end{equation}
where the coefficients $A_{m,n} = \int F(x,y)*\phi_{L,m}(x)\phi_{L,n}(y) dxdy$ of the expansion give a coarsest level-$L$ approximation of the field, and the wavelet coefficients $D_{p,l,m,n} = \int F(x,y)*\Psi^p_{l,m,n}(x,y)dxdy$ retain information on the finer-level details.

\section{Iterative Wavelet Filtering}
\label{sec:iterative}

It has been suggested that turbulence in fluid flows may be characterized by the presence of a strong incoherent portion \cite{farge:1992}. The incoherent background may be modeled as a stochastic forcing term \cite{schneider:2010} on a collection of coherent structures. Wavelet techniques have been employed to analyze direct numerical simulations \cite{farge:1999} as well as serving as a basis for coherent vortex simulations \cite{schneider:2006}.

We apply below an iterative wavelet filtering method \cite{donoho:1994,farge:1999,azzalini:2005} to the current density from numerical solutions of turbulent plasma flows. The iteration procedure determines an optimal cut-off threshold for the $N$ wavelet coefficients $\{A_{m,n},D_{p,l,m,n}\}$, which we refer to generically as $\{C_I\}$. Coefficients $C_I$ with modulus below the threshold (which is user-defined by a multiplicative factor $\alpha$ as outlined below) are classified as part of a background of incoherent noise. The coefficients with modulus above the threshold contribute to the coherent features of the flow. The incoherent noise is assumed to be additive, Gaussian, and white \cite{schneider:2006,okamoto:2007}. The method proceeds as follows:

\begin{enumerate}
\item A choice is made of a multiplicative factor $\alpha$, the number of levels in the wavelet decomposition $L$, and the wavelet basis. The number of levels $L$ is chosen so that $2^L\sim min(N_x,N_y)$ where $N_x$ and $N_y$ are the number of computational grid points in each direction of the domain. Especially for particle simulations with intrinsic statistical noise, a value of $\alpha > 1$ is necessary to capture features of the turbulent flow rather than grid-scale noise. (See the Appendix for a discussion of extracting particle noise using the wavelet filter.)
\item The current density is expressed as a sum over an ortho-normal wavelet basis with $N$ coefficients $\{C_I\} = \{A_{m,n},D_{p,l,m,n}\}$.
\item A threshold $\epsilon = \alpha\sqrt{Var(\{C_I\})log(N)}$ is initialized based on the variance of the set of coefficients $\{C_I\}$. 
\item The incoherent (or noise) portion of the current density is defined by the coefficients $\hat{C_I}$, where $\hat{C_I} = C_I$ if $|C_I| < \epsilon$ and the remaining coefficients with $|{C_I}|>\epsilon$ are set to zero.
\item A new threshold $\epsilon = \alpha\sqrt{Var(\hat{C_I})log(N)}$ is computed based on the extracted noise.
\item Steps 4 and 5 are repeated until the threshold $\epsilon$ varies less than 5\% over an iteration. In practice, the method typically converges to this tolerance after 2-4 iterations.
\item Finally, the coherent part of the current density is reconstructed from the wavelet coefficients $C_I$ with $|C_I|>\epsilon$. The incoherent current density is obtained by subtracting the coherent one from the total one pointwise (or equivalently by inverse wavelet transform from the weak wavelet coefficients with $|C_I|\le\epsilon$). 
\end{enumerate}

\begin{figure}
\includegraphics[width = 0.8\textwidth]{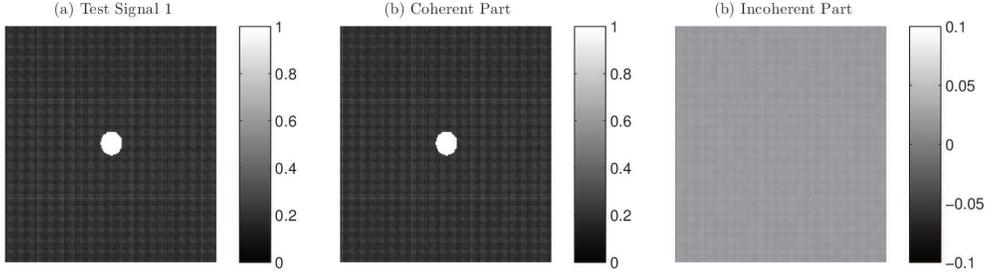}
\caption{The iterative wavelet filter is used to decompose Test Signal 1 from Fig.~\ref{fig:coherent}(a) into coherent and incoherent parts. The incoherent part is very small. \label{fig:wcoh}}
\end{figure}

\begin{figure}
\includegraphics[width = 0.8\textwidth]{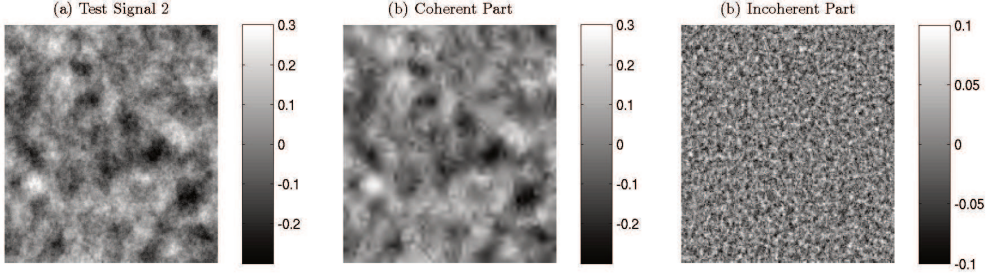}
\caption{The iterative wavelet filter is used to decompose Test Signal 2 from Fig.~\ref{fig:coherent}(b) into coherent and incoherent parts. The coherent part is represented by only $\sim0.07\%$ of the wavelet coefficients. \label{fig:winc}}
\end{figure}

To illustrate the effect of the iterative wavelet filter, we apply it to the two test signals of Fig.~\ref{fig:coherent}(a-b). The decomposition into coherent and incoherent parts as defined by the wavelet filter are plotted in Figs.~\ref{fig:wcoh} and~\ref{fig:winc}, where for each case we set the parameters $\alpha=10$ and $L=8$ and used the coiflet-18 basis \cite{daubechies:1993}. The main conclusions do not depend on the choice of wavelet basis. Nevertheless, it is useful for turbulent flows to choose a basis with a large number of vanishing moments (the coiflet-18 wavelets have six vanishing moments), which tends to cancel the wavelet coefficients in smooth regions free of discontinuities and high-order derivatives \cite{farge:1992}. For the localized, coherent Test Signal 1 in Fig.~\ref{fig:wcoh}, the method finds an extremely small incoherent noise part. For Test Signal 2 in Fig.~\ref{fig:winc}, a large background noise is extracted. Note that the coherent portion of the signal in Fig.~\ref{fig:winc}(b) is reconstructed from only $\sim0.07\%$ of the wavelet coefficients even though it contains over $99\%$ of the "energy" $\sum_n F_n^2$ of the signal. This ability to capture a large portion of a signal with a small number of coefficients explains the wide-spread use of wavelets for digital signal compression \cite{chang:2000}.

\section{Simulation Set-Up}
\label{sec:sim}

To study the transition to turbulence of a magnetized plasma flow, we consider 2D simulations of Kelvin--Helmholtz unstable flow-shear layers using codes employing three different models: (1) fully kinetic particle-in-cell modeling using the code VPIC \cite{bowers:2008}, 
(2) hybrid kinetic ion/fluid electron modeling using the hybrid PIC code H3D \cite{karimabadi:2011}, and (3) Hall-MHD modeling using the PIXIE3D code \cite{chacon:2008,stanier:2017}. 

The first simulation is a fully kinetic simulation performed with the code VPIC that was analyzed in Ref.~\cite{karimabadi:2013}, and it is referred to here as "VPIC A". A plasma of uniform density and magnetic field (mainly out of the simulation plane, but with a $5\%$ component added in the initial flow direction). The velocity distribution of each species $s=i,e$ (ion and electron) is a drifting Maxwellian with uniform temperature $T_s$ giving a species $\beta_s=0.05$, and with a drift speed $U_y=U_0\tanh(x/L)$. The shear layer half-thickness $L=4d_i$ and the flow speed $U_0=10V_A$, where $V_A$ is the Alfv\'en speed. Periodic boundary conditions are imposed in $y$, while the boundary conditions at $x=0$ and $x=L_x$ are conducting for electromagnetic fields and reflecting for particles. Other numerical parameters are $m_i/m_e=100$ and $\omega_{pe}/\omega_{ce}=2$. Further details are found in the table below and Ref.~\cite{karimabadi:2013}. 

One of the serious limitations of PIC codes is statistical noise associated with using a relatively small number of computational particles to sample the distribution function. The large-scale simulation "VPIC A" used 150 particles per cell per species, which is representative of many other simulations in the literature. In order to understand the sensitivity of our results to noise, we also analysed a VPIC simulation with 10,000 particles per cell, which will be referred to as "VPIC B" simulation. The parameters are largely analogous to the "VPIC A" case, except for a smaller spatial extent of the simulation domain, a smaller initial width of the transition layer $L=0.5d_i$ (see Table~\ref{tab:runs}), and higher plasma beta of $\beta_s=0.15$.

Two additional simulations using different plasma models are next considered. In each case, the system is doubly periodic in a domain of size $L_x\times L_y=7.5\pi\times5\pi d_i$, where $d_i$ is the ion inertial skin depth based on the uniform density $n_0$. The initial magnetic field is uniform and mainly out of the simulation plane, $B_z=B_0$, with a small additional in-plane component $B_y=0.05B_0$. Two flow-shear layers are given with velocity profiles:
\begin{equation}
\frac{v_y(x)}{v_0} = \tanh\left(\frac{x-L_x/4}{\lambda}\right) - \tanh\left(\frac{x-3L_x/4}{\lambda}\right) - 1,
\end{equation} 
where $v_0 = 0.5 v_A = 0.5 B_0/\sqrt{4\pi n_0m_i}$ and the scale length $\lambda = 0.5 d_i$. A motional electric field ${\bf{E}}(x) = -{\bf{v}}(x)\times{\bf{B}}$ is also included included. For each simulation, we focus on only one of the shear-flow layers, in particular whichever layer transitions to a turbulent state fastest. 

\begin{table}
\caption{Parameters of numerical simulations. VPIC is a fully kinetic particle-in-cell code \cite{bowers:2008}, H3D is a hybrid kinetic ion/fluid electron code \cite{karimabadi:2014}, and PIXIE3D is a Hall MHD code \cite{chacon:2016}.}
\begin{center}
\begin{tabular}{|c|c|c|c|c|c|}
\hline
Code 		& System Size			&  Cells			&  Particles 		& $\alpha$ 	&Other Numerical \\
Run			& ($d_i$)   			& 	      			&					& 			&Parameters      \\
\hline
VPIC A 		&$100 \times 50$ 		&$16384\times8192$ 	&$4\times10^{10}$ 	&3			&$m_i/m_e=100$	 \\
VPIC B 		&$23.6 \times 15.7$ 	&$3328\times2560$ 	&$8.5\times10^{10}$	&5			&$m_i/m_e=100$   \\
H3D 		&$23.6 \times 15.7$ 	&$1536\times1024$ 	&$3\times10^{9}$ 	&3&			$\eta=10^{-7}$   \\
PIXIE3D 	&$23.6 \times 15.7$ 	&$2048\times1024$ 	& N/A 	&2	&$\mu_i=10^{-4}$,$\chi_e=10^{-4}$,$\mu_e=10^{-6}$ \\
\hline
\end{tabular}
\end{center}
\label{tab:runs}
\end{table}

For the Hall-MHD simulation, additional parameters included an ion viscosity $\mu_i=10^{-4}$, a heat conductivity of  $\chi_e = 10^{-4}$, and an electron viscosity (hyper-resistivity) $\mu_e = 10^{-6}$. The latter value was chosen to set a sub-$d_i$ dissipation scale for current-layers, to prevent them from thinning down to grid-scale. Time advance used the BDF-2 method with a time-step $\Delta t = 10^{-3}$. The Hall-MHD equations are spatially discretized on a cell-centered mesh using central differences \cite{chacon:2004}, apart from the advection terms that were treated with the monotonicity preserving SMART algorithm~\cite{gaskell:1988}. The required spatial resolution was found from a grid-convergence study, where the chosen value of $2048\times1024$ cells was found to give a converged value of the peak magnetic energy just prior to the transition to turbulence at $t\Omega_{ci} =40$.

The hybrid-PIC simulation uses an approximation where ions are treated kinetically, while electrons are represented as a massless fluid. The simulation analyzed here was conducted using a version of the H3D code~\cite{karimabadi:2011} optimized for turbulence simulations~\cite{podesta:2017}.

\section{Measuring the Transition to Turbulence}
\label{sec:transit}

Our goal is to test the iterative wavelet filtering method on each of the three types of plasma simulation to determine if the wavelet analysis is capable of identifying the onset of turbulence. Wavelet techniques have been used previously for analyzing the transitional regime to turbulence in a boundary layer of a rotating disk in hydrodynamic turbulence and to estimate the transitional Reynolds number \cite{moret-bailly:1991}. While there are differences in the details of the current sheets and flows between the various plasma simulation models, we do not attempt to characterize these differences here. Indeed, a positive feature of wavelet analysis is that it does not pre-suppose a model for the coherent structures that arise in the turbulent flow.

\subsection{Large Fully Kinetic Run}

\begin{figure}
\includegraphics[width = 0.8\textwidth]{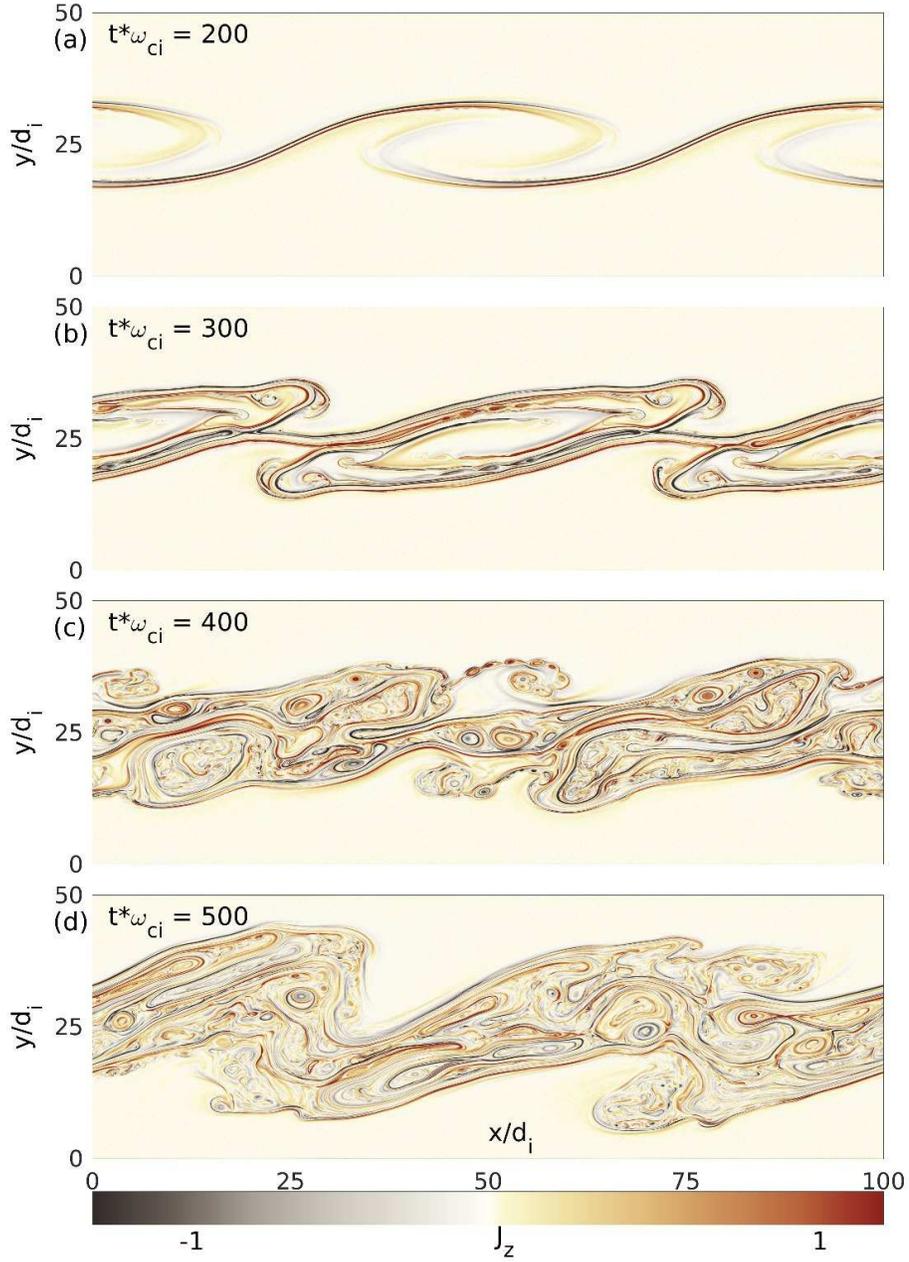}
\caption{Out of plane current density $J_z$ from VPIC A run (also see Ref.~\cite{karimabadi:2013}) at four different times. \label{fig:khpop}}
\end{figure}

We first test the wavelet turbulence diagnostic on the large 2D full kinetic Kelvin-Helmholtz simulation ("VPIC A") previously analyzed in Ref.~\cite{karimabadi:2013}. The out-of-plane current density from the simulation is plotted at four different times in Fig.~\ref{fig:khpop}. Current sheets form as the in-plane magnetic field is advected with the shear flow, highlighting the main large-scale Kelvin-Helmholtz vortices that have nearly saturated in magnitude at time $t=200/\omega_{ci}$ in Fig.~\ref{fig:khpop}(a). An important process for transferring energy to smaller scales in a cascade is secondary tearing \cite{karimabadi:2013}, which breaks the developing current sheets into a series of magnetic islands or plasmoids \cite{comisso:2016}. The development of plasmoids is a primary trigger in this system for the transition to turbulence. A chain of secondary magnetic islands is visible in Fig.~\ref{fig:khpop}(b) at time $t=300/\omega_{ci}$. By time $t=400/\omega_{ci}$ in Fig.~\ref{fig:khpop}(c), a number of current sheets and magnetic islands across a range of scales have developed. While this secondary magnetic reconnection process is sufficient on its own to generate turbulence \cite{huang:2016,daughton:2011}, the nonlinear development depends on the details of the global system. In our case, the imposed background velocity shear continues to couple to the magnetic islands, and it forces both island merging and additional tearing. 

\begin{figure}
\includegraphics[width = 0.95\textwidth]{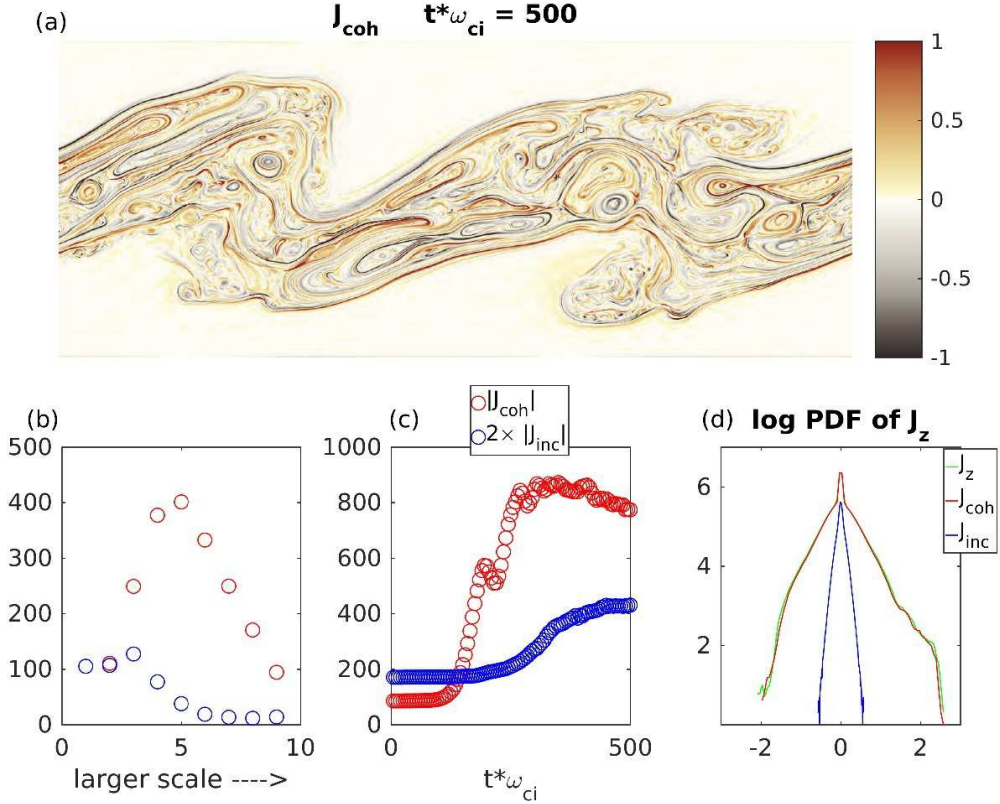}
\caption{Diagnostics applied to VPIC A simulation. (a) Profiles of the out-of-plane current density $J_z$ at the end of the simulation. (b) The norms ($\sqrt{\sum J^2}$) of the coherent (red) and incoherent (blue) portions of the current density separated by scale in the wavelet multi-resolution decomposition. Spatial scale increases to the right.  (c) Total norm of the coherent (red) and incoherent (blue) parts over time. The transition to turbulence at $t\sim$250---350$/\omega_{ci}$ is marked by the increase of the incoherent background. (d) Log of the probability distribution function of values of $J_z$ over the simulation domain. The coherent portion (red) has an extended super-Gaussian tail at larger values. (Multimedia view online.) \label{fig:khpop2}}
\end{figure}

We apply the iterative wavelet diagnostic to measure this transition to a turbulent state. The current density is de-composed into coherent and incoherent portions as defined by the wavelet threshold method of Section~\ref{sec:iterative}. The wavelet decomposition here uses 10 wavelet levels, spanning the grid scale to nearly the global scale of the 2D run. The norms $|J| = \sqrt{\sum J^2}$ of the coherent and the incoherent portion at the end of the simulation at time $t=500/\omega_{ci}$ are plotted in Fig.~\ref{fig:khpop2}(b). While the incoherent portion contains a contribution from grid-scale numerical noise associated with particle methods (see the Appendix), it acquires additional energy particularly at micro- or meso-scales (peaked at level 3) as the flow transitions to turbulence. (See movie version of Fig.~\ref{fig:khpop2}.)

The growth of the incoherent part as the shear layer transitions to turbulence is apparent in Fig.~\ref{fig:khpop2}(c). Here, the norms of both the coherent (red) and incoherent (blue) portions are plotted over time. The coherent portion grows rapidly as the global-scale Kelvin-Helmholtz instability with a growth rate of $\gamma\sim80\omega_{ci}$ develops, noticeably increasing at $t\sim150/\omega_{ci}$. During this period, the large, coherent, global-scale vortices form. As secondary tearing and other processes cause a cascade down to smaller scales, the flow becomes turbulent. At this stage, the incoherent portion grows in size, particularly around $t\sim$250---350$/\omega_{ci}$. We identify this growth of the incoherent portion as the marker of a transition to turbulence. The incoherent part has a probability distribution function (PDF) that is approximately Gaussian [see Fig.~\ref{fig:khpop2}(d)], while the coherent part includes a tail of stronger, intermittent structures skewed to larger values.  

\subsection{Comparison of Different Models}

\begin{figure}
\includegraphics[width = 0.95\textwidth]{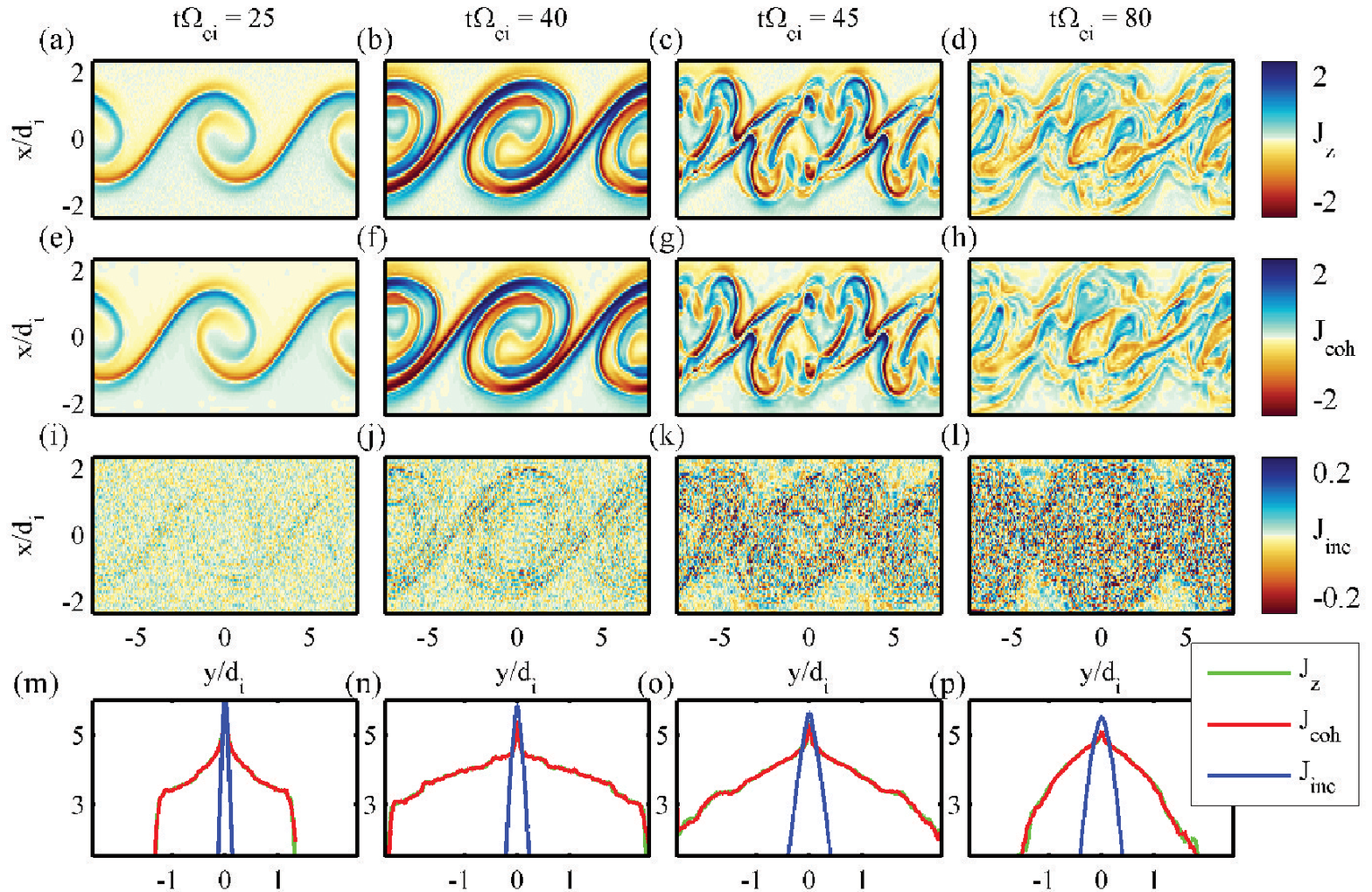}
\caption{Fully kinetic PIC results from VPIC B simulation. (a-d) Profiles of the out-of-plane current density $J_z$ over the course of a VPIC simulation. (e-h) The coherent part $J_{coh}$ of $J_z$ extracted through the iterative wavelet filtering. (i-l) The residual incoherent part $J_{inc}=J_z-J_{coh}$. (m-p) PDFs of the total out-of-plane current density $J_z$ (green), the coherent portion extracted by the wavelet method $J_{coh}$ (red), and the incoherent portion $J_{inc}$ (blue). \label{fig:jz}}
\end{figure}

In this section, we apply the same iterative wavelet method to the three simulations of varying type that modeled a smaller shear flow layer. The out-of-plane current density $J_z$ is plotted in Fig.~\ref{fig:jz}(a-d) at four time slices over the course of the high-resolution VPIC simulation with 10,000 particles per cell and $3328\times2560$ cells. The coherent part of the current $J_{coh}$ extracted through the wavelet method yields the profiles in Figs.~\ref{fig:jz}(e-h). The percentage of wavelet coefficients required to reconstruct the coherent portion ranges from $\sim0.06\%$ at $t\Omega_{ci}=20$ to $\sim0.13\%$ at $t\Omega_{ci}=80$. Nevertheless, this small fraction of coefficients contains $99\%$ of the ``energy" (defined as $\sum J_z^2$) at $t\Omega_{ci}=20$ and $96\%$ at $t\Omega_{ci}=80$. 

\begin{figure}
\includegraphics[width = 0.95\textwidth]{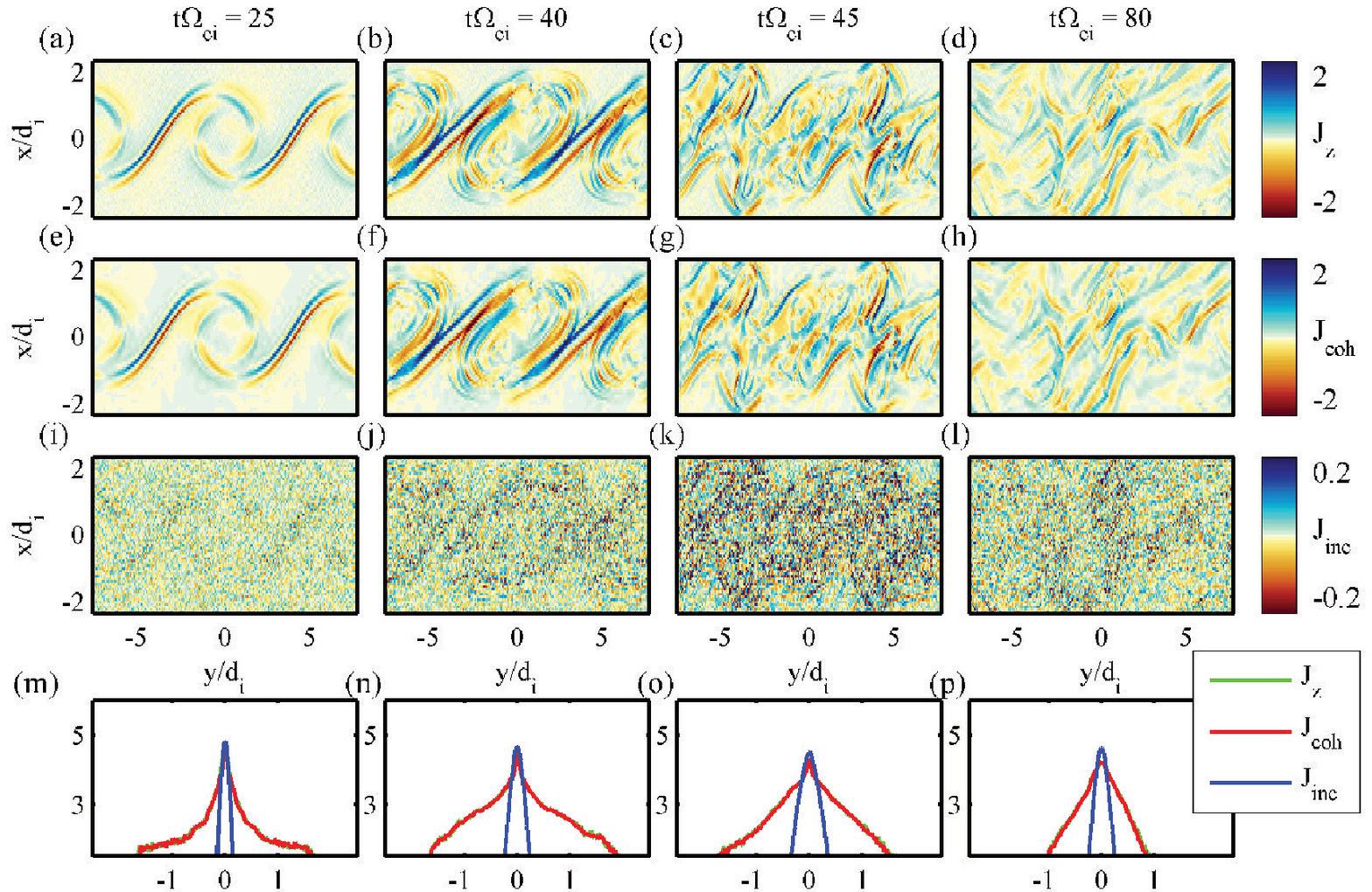}
\caption{Hybrid (kinetic ion/fluid electron) PIC simulation results. Similar plots as in Fig.~\ref{fig:jz}. \label{fig:jzh3d}}
\end{figure}

\begin{figure}
\includegraphics[width = 0.95\textwidth]{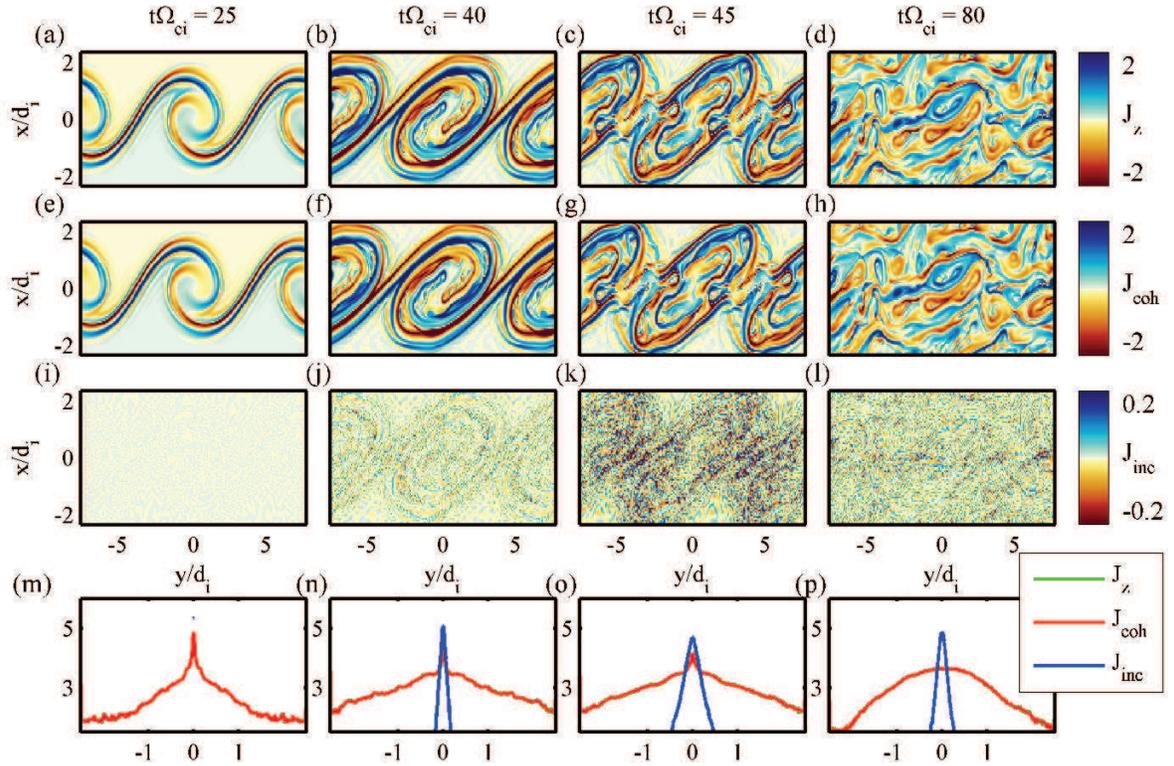}
\caption{Hall MHD simulation results. Similar plots as in Fig.~\ref{fig:jz}. \label{fig:jzhmhd}}
\end{figure}

The logarithm of the PDF of values of $J_z$ over the entire simulation domain is plotted for each time slice in Figs.~\ref{fig:jz}(m-p). The wavelet filtering technique extracts a large coherent portion of each current density profile. The coherent piece contains slowly decaying PDFs, which are approximately power laws for a range of values. The coherent portion of the current density thus includes intermittent current sheets, which have been identified as key sites of dissipation in kinetic turbulence \cite{wan:2012,karimabadi:2013}. The incoherent portion (blue curves) is nearly Gaussian noise, corresponding to parabolic profiles in the logarithmic plots. Similar plots from each of the simulations are shown in Figs.~\ref{fig:jzh3d}-\ref{fig:jzhmhd}.

\begin{figure}
\includegraphics[width = 0.7\textwidth]{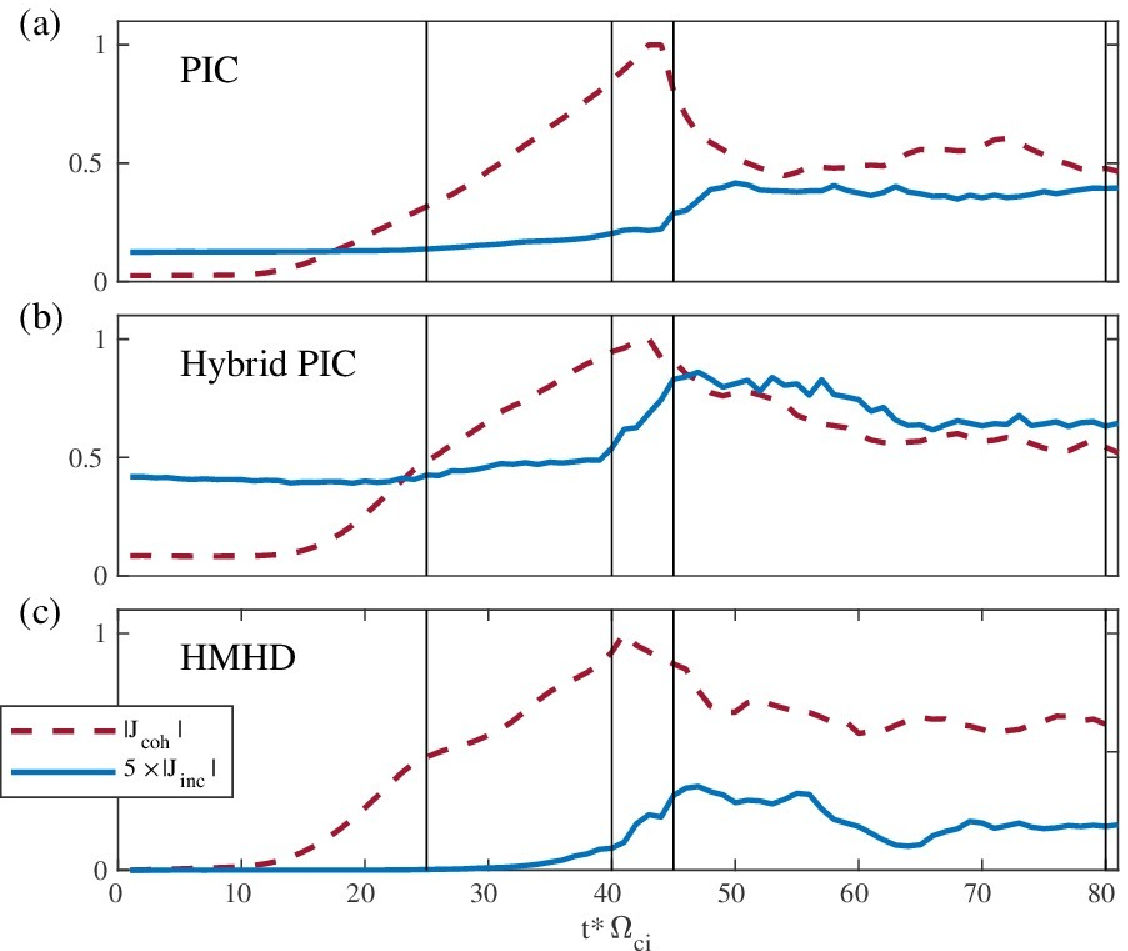}
\caption{The norm of the current density, $|J| = \sqrt{\int |J(x,y)|^2 dx dy}$, over the simulation domain of the total out-of-plane current $J_z$ (black) and the incoherent portion $J_{inc}$ (red, scaled by a factor of 5) are plotted over time for (a) the fully kinetic PIC simulation, (b) the hybrid PIC simulation, and (c) the Hall MHD simulations. Each is normalized to the maximum $|J_z|$. The vertical lines indicate the time steps used in Figs.~\ref{fig:jz}-\ref{fig:jzhmhd}. The incoherent part increases and then saturates as the dynamics become turbulent. \label{fig:jnorms}}
\end{figure}

As for the larger fully kinetic run, to study the transition to turbulence as the initial laminar Kelvin--Helmholtz vortices break apart into smaller structures and generate dissipation-scale current sheets, we plot the norm $|J| = \sqrt{\sum J^2}$ of the total out-of-plane current density $|J_z|$ and the incoherent part $|J_{inc}|$ over time for each simulation in Fig.~\ref{fig:jnorms}. In each case, the total current density norm $|J_z|$ increases as the Kelvin--Helmholtz vortex forms. Even when the vortex reaches a non-linear state near the maximum of $|J_z|$, the incoherent part $|J_{inc}|$ remains small. The incoherent part $|J_{inc}|$ then undergoes a relatively rapid increase in magnitude as the vortex breaks apart through turbulent motions and generates current sheets over a range of length scales. Again, we associate this rapid increase and subsequent saturation of the incoherent part $|J_{inc}|$ with the onset of turbulence.

\begin{figure}
\includegraphics[width = 0.8\textwidth]{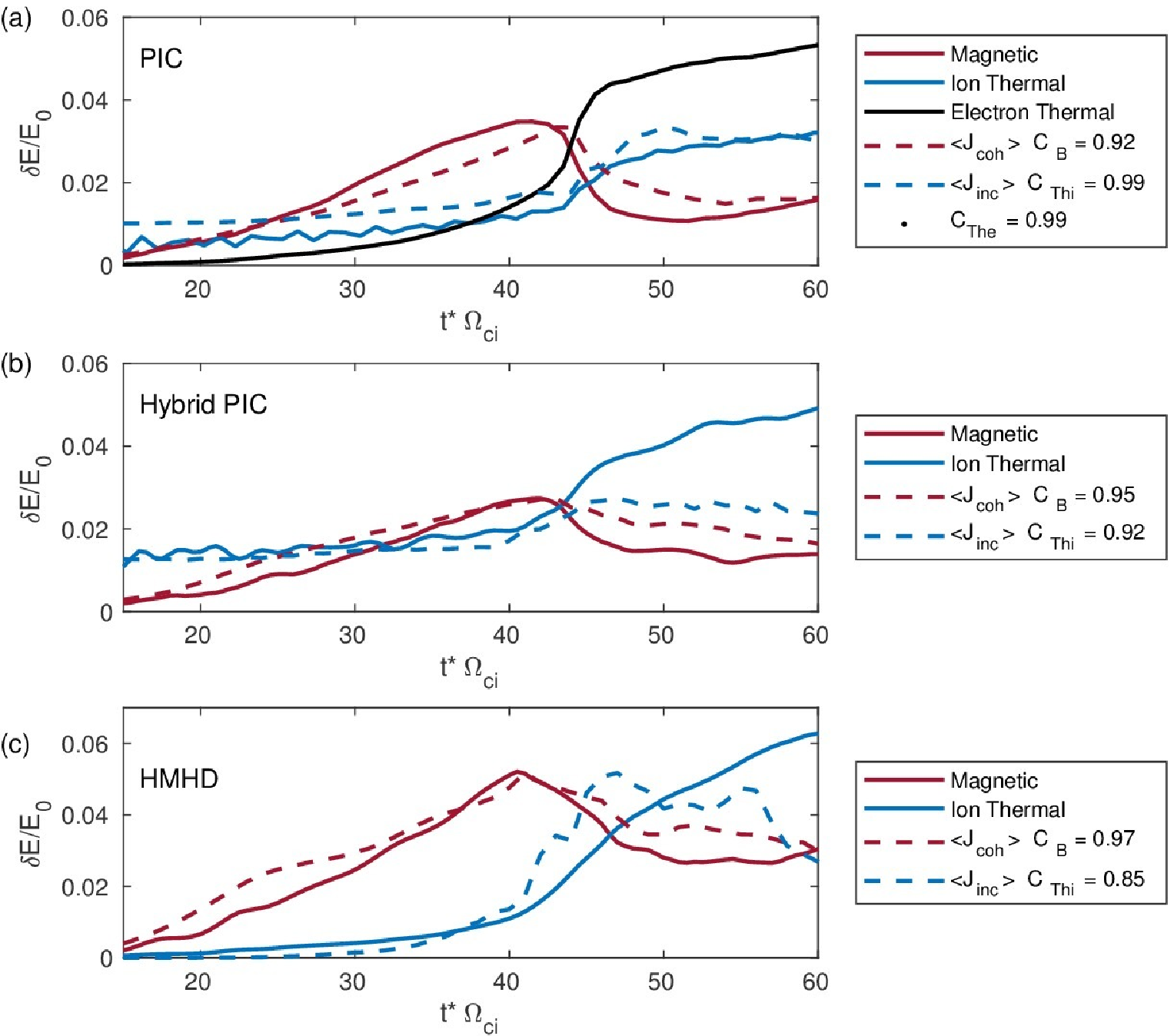}
\caption{The change in magnetic energy and thermal energy over an interval of each simulation (summed over the simulation domain and normalized to the initial total ion kinetic energy associated with the shear flow). The norms of the coherent $J_{coh}$ and incoherent $J_{inc}$ out-of-plane current density are over-plotted, normalized by the maximum in magnetic energy change. Note that the magnetic energy change is highly correlated with the coherent current density. The Pearson correlation coefficients $C_B\in[-1,1]$ between magnetic energy change and $|J_{coh}|$ are given in the figure legend. Likewise, the incoherent current density correlates with the increase in ion thermal energy as quantified by the coefficients $C_{Thi}$ (and $C_{The}$ based on electron thermal energy in the fully kinetic run) between thermal energy and $|J_{inc}|$.   \label{fig:energies}}
\end{figure}

Interestingly, the growth in the coherent $J_{coh}$ and incoherent $J_{inc}$ portions of the current density [plotted in Fig.~\ref{fig:jnorms}] correlate  with the transfer of energy from the ion flow to magnetic energy and plasma thermal energy. As the initial shear flow carries the in-plane field and generates current sheets around the Kelvin-Helmoltz vortices, kinetic energy is transferred to the magnetic field. This is illustrated in Fig.~\ref{fig:energies}, where the change in total magnetic energy in the system is plotted in red along with the coherent portion of the current density $|J_{coh}|$ (red dashed curves). Because this initial rise in magnetic energy is associated with the large-scale coherent vortices, the two red curves are highly correlated. In particular, the Pearson correlation coefficient $C_B\in[-1,1]$  between the magnetic energy and $|J_{coh}|$ for each simulation is $>0.9$. In contrast, the correlation coefficient between the magnetic energy and $|J_{inc}|$ in VPIC Run B, for example, is -0.03. 

Once strong turbulence develops, energy is converted into thermal energy, and the plasma temperature increases. For this reason, the incoherent current density $|J_{inc}|$, which displays an uptick when turbulence develops, is highly correlated with the plasma thermal energy [see blue curves in Fig.~\ref{fig:energies}]. Indeed, for the fully kinetic simulation, the correlation coefficient $C_{Thi}$ between $|J_{inc}|$ and the ion thermal energy, and the coefficient $C_{The}$ between $|J_{inc}|$ and the electron thermal energy are both 0.99. The hybrid model and Hall-MHD models also show strong correlation ($>0.85$) between ion thermal energy and $|J_{inc}|$.

\subsection{Wavelet Technique Versus Other Diagnostics of Turbulence}

In this section, we evaluate the utility of three diagnostic methods that are typically used in turbulence by tracking their behavior from the laminar to the fully developed turbulent stage. The three methods are Fourier power spectra, structure function, and finite time Lyapunov exponent (FTLE).

\begin{figure}
\includegraphics[width = 0.6\textwidth]{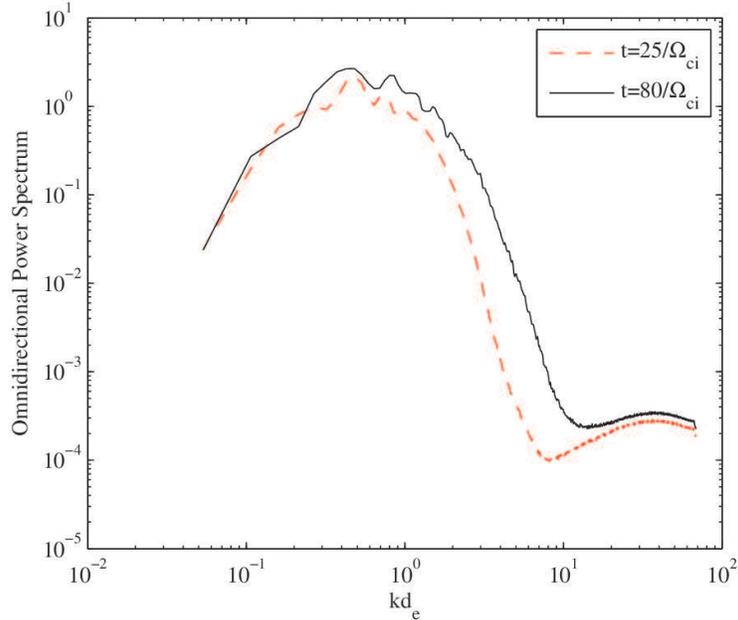}
\caption{Fourier spectra of $J_z$ from the fully kinetic VPIC simulation at early time $t=25/\Omega_{ci}$ (red) before the onset of turbulence and late time $t=80/\Omega_{ci}$ (black) after turbulence has fully developed.\label{fig:jspect}}
\end{figure}

We start by showing the Fourier power spectra of the current density from the fully kinetic simulation at early ($t=25/\Omega_{ci}$)  and late ($t=80/\Omega_{ci}$) time in Fig.~\ref{fig:jspect}. The late-time, turbulent spectrum (black) does have a stronger signal at somewhat larger $k$ and has overall greater energy than the early-time, laminar spectrum (red). Nevertheless, the two spectra have very similar shapes and slopes, with no clear indication of a transition to a turbulent state.

\begin{figure}
\includegraphics[width = 0.6\textwidth]{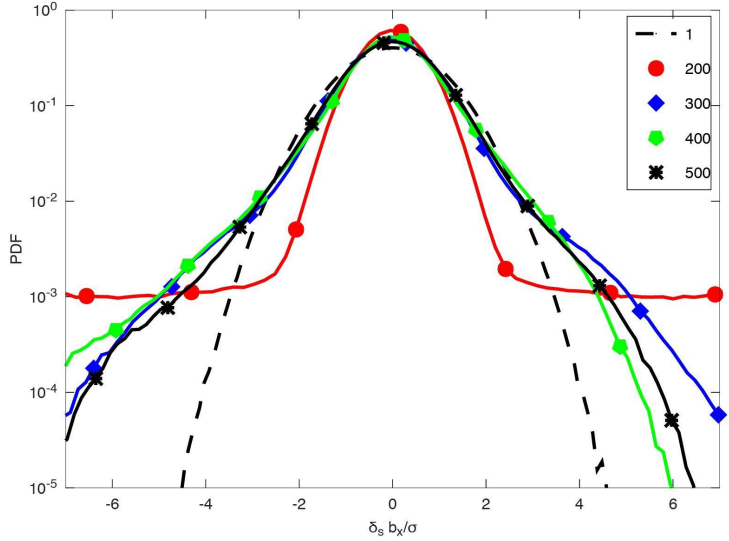}
\caption{Normalized PDF of magnetic field increments in VPIC case A at $t\Omega_{ci} = 1,200,300,400,500$. Here $\delta_s b_x = b_x(y
+ s) - b_x(y)$ and $\sigma$ is its standard deviation. For all curves $s=1d_e$, which is the scale where the strongest deviation from Gaussian distribution is observed in the fully developed turbulence.  \label{fig:nongauss}}
\end{figure}

Many statistical techniques beyond power spectra have been developed for characterizing turbulence. For example, statistics of field increments, often characterized by structure functions, can reveal information about energy fluxes, and degree of intermittency. Fig.~\ref{fig:nongauss} shows the normalized PDFs of magnetic field increments in one of the simulations analyzed here (VPIC case A). Deviations from Gaussian PDF at a given scale indicate intermittency at that scale. As is evident from this figure, there are signatures of non-Gaussianity even in the early stages of developing turbulence (e.g., $t\Omega_{ci} = 300$). Thus, it is difficult to unequivocally distinguish relatively early stages of developing turbulence from the fully developed turbulence with this technique.

\begin{figure}
\includegraphics[width = 0.9\textwidth]{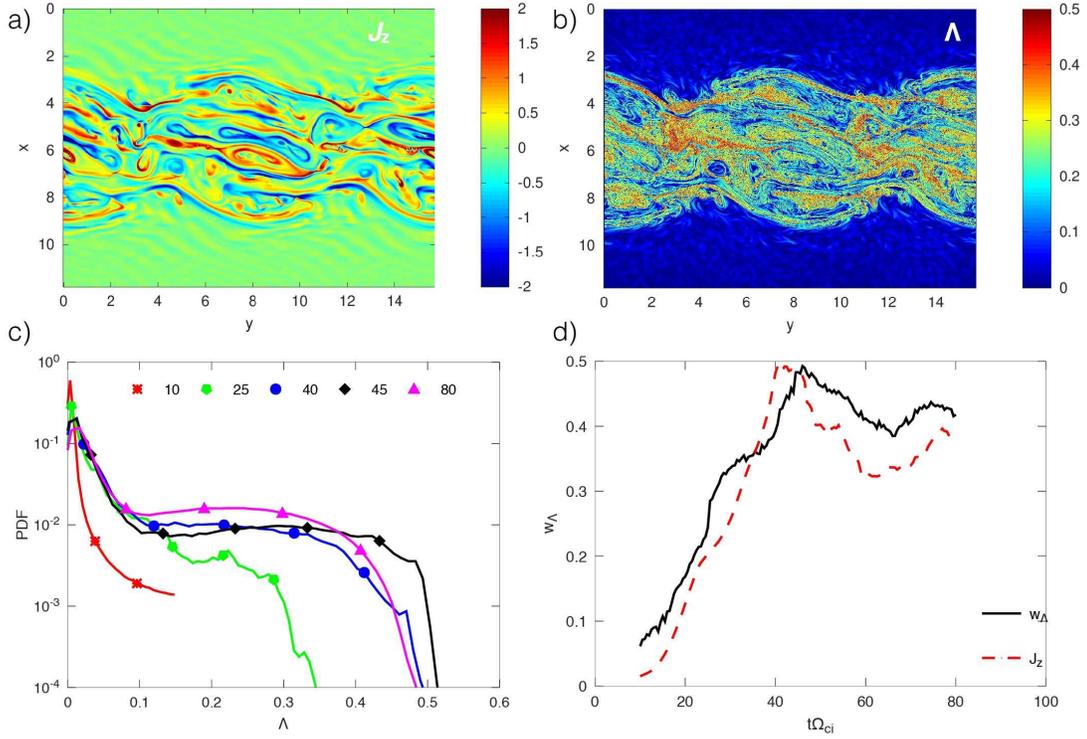}
\caption{FTLE analysis of the Hall-MHD simulation: a) $z$-component of the current density $J_z$; b) FTLE ($\Lambda$) field computed in backward time starting from time $t\Omega_{ci}=80$; c) PDF of FTLE values for several starting times; d) evolution of the PDF width at the value $3 \times 10^{-3}$ in time with a re-scaled root-mean-squared value of $J_z$ overlayed for reference. \label{fig:FTLE}}
\end{figure}

The wavelet decomposition allows one to draw a distinction between spatially coherent and incoherent parts of a turbulent field. But coherency of a certain field could also be analyzed in the temporal sense. Here we consider an analysis based on computation of FTLE. Indeed, local maxima of FTLE computed in backward (forward) time may indicate attractive (repelling) Lagrangian coherent structures~\cite[e.g.][]{haller:2011}. We focus on the noise-free Hall-MHD simulation and compute FTLE of the electron flow by integrating fluid trajectories originating from neighboring grid points over a time interval of $T\Omega_{ci}=10$ in backward time. Panel b) in Fig.~\ref{fig:FTLE} shows the resulting FTLE values. While we do not attempt to extract FTLE ridges and thus accurately determine the (attractive) coherent structures here, the visual inspection reveals an abundance of localized regions with  high values of FTLE. Furthermore, these regions often appear in the vicinity of enhanced current density, which is shown in Panel a) of Fig.~\ref{fig:FTLE}. As the turbulence develops, distribution of FTLE values evolves and becomes significantly wider, as is illustrated in Panels c) and d) of Fig.~\ref{fig:FTLE}. The width of the FTLE PDF shows a strong correlation with the temporal evolution of current density. This provides confidence in FTLE as a tool for study of coherent structures in turbulence. However, unlike the incoherent wavelet component that was mainly flat but showed a rapid rise close to the onset of turbulence, the width of FTLE PDF rises almost in lockstep with the growth of current density during the initial phase of the KH instability, and it continues to rise until it reaches an overshoot point. It then settles down to an asymptotic state during the fully developed phase of turbulence. We  conclude that while FTLE analysis, at least in the way that we have used it here, gives a certain indication for development of turbulence, it alone cannot be used to unequivocally distinguish a fully turbulent state from that of developing turbulence.  It should also be noted that unlike the wavelet method, FTLE analysis does not suggest any quantitative value to indicate whether the system is in a fully developed turbulence stage.

\section{Turbulence in a Global Magnetosphere Simulation}
\label{sec:global}

In this section, we apply the above wavelet techniques to analyze turbulence in a 2D hybrid global magnetosphere model \cite{karimabadi:2014}. The model consists of a fixed dipolar magnetic field (enclosed in a conducting spherical "planet") and a solar wind entering the simulation from the left with a specified Alfvenic Mach number. A bow shock forms where the flow collides with the planetary magnetic field. Behind the bow shock is the magnetosheath, which is a shocked solar wind plasma separated from the planetary dipole field by an inner boundary layer called the magnetopause. While the magnetosheath is mainly laminar in the quasi-perpendicular region of the bow shock, it is highly turbulent in the quasi-parallel region of the bow shock.
This turbulence is generated by ion kinetic effects and would be absent in magnetized fluid models of the bow shock. Ion kinetic effects lead to the formation of the ion foreshock, an extended region upstream of the bow shock driven by beams of ions reflected from the shock. The foreshock instabilities are strongest and most spatially extended along magnetic field lines that are quasi-parallel to the shock normal. The foreshock fluctuations steepen and develop into turbulence and coherent jets when they are advected into the magnetosheath \cite{karimabadi:2014}.

\begin{figure}
\includegraphics[width = 0.9\textwidth]{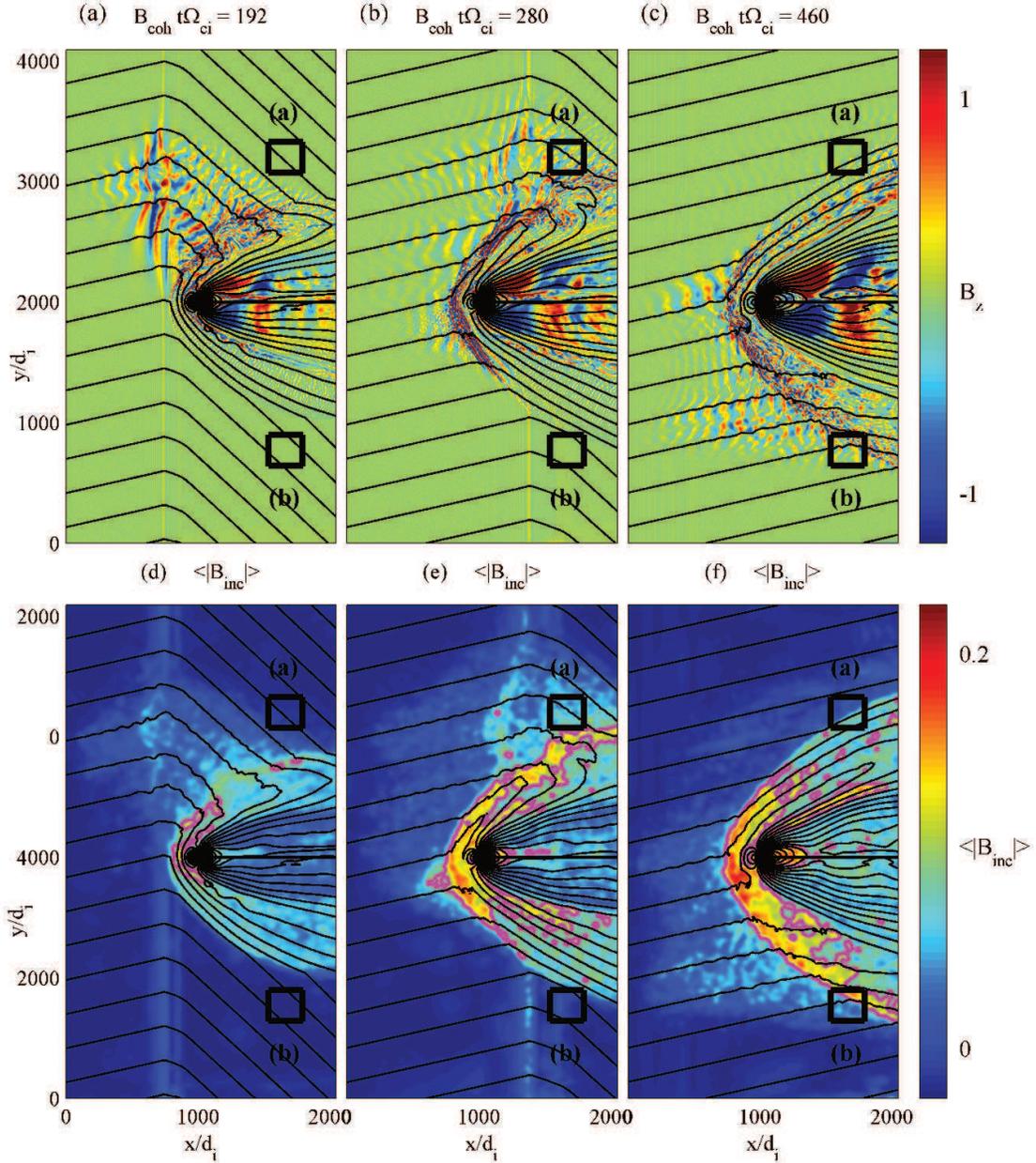}
\caption{(a-c) The coherent part of the out-of-plane magnetic field $B_Z$ from the global hybrid magnetic field run. In (d-f), the absolute value of the incoherent part is smoothed by convolution with a Gaussian kernel of variance $\sim50$ cells. Turbulence is associated with regions containing a large incoherent signal, with the magenta contour indicating a level of 0.1 $B_0$. The boxes indicate the regions where localized turbulence is analyzed in Fig.~\ref{fig:boxnorms}. (See multimedia view online.) \label{fig:globalinc}}
\end{figure}

We conduct our analysis for Run 1 in Ref.~\cite{karimabadi:2014} with Alfvenic Mach number of 8, domain size $N_z\times N_x=8192\times 2048$ cells, a cell resolution of $\Delta x=1d_i$ ($d_i$ is the ion inertial length), and with 200 particles per cell. This run is of particular interest since the direction of the interplanetary magnetic field reverses in time, launching a rotational discontinuity in the solar wind. The corresponding sharp rotation of the magnetic field direction is visible in Figs.~\ref{fig:globalinc}(a-b). (Note that the domain has been truncated in the $z$ direction for the plots.) As the discontinuity crosses the planetary magnetosphere, the region where the magnetic field is quasi-parallel to the shock normal changes. As a result, the strongest foreshock and magnetosheath fluctuations move from the northern hemisphere to the southern hemisphere (compare Figs.~\ref{fig:globalinc}(a) and (c) for example).  We consider here whether the wavelet analysis techniques offer a means of quantifiying this localized shift of the turbulent flow features in this highly inhomogeneous system.

In Figs.~\ref{fig:globalinc}(a-c), we plot the coherent part $B_{coh}$ of the fluctuations of the magnetic field component $B_z$ out of the simulation plane, extracted using the iterative wavelet filter at three different times over the course of the simulation. The incoherent part $B_{inc}=B_z-B_{coh}$  is used to locate regions of turbulence. To yield a measure of turbulence associated with small extended regions of the simulation (rather than a purely local measure), the absolute value of the incoherent part $B_{inc}$ is convolved with a Gaussian filter of width of $\sim 80$ cells. This smoothed average value $<|B_{inc}|>$ is also plotted in Figs.~\ref{fig:globalinc}(d-f). Following the results of the analysis of Kelvin--Helmholtz unstable flows of Sec.~\ref{sec:transit}, we identify the turbulent regions as those with a large incoherent signal. Selecting a threshold of $<|B_{inc}|>=0.1B_0$ produces the magenta contours plotted in Figs.~\ref{fig:globalinc}(d)-(f), and these contours thus contain the turbulent regions as defined by the wavelet analysis.

Fig.~\ref{fig:globalinc}(a) shows that the coherent component clearly captures the formation of wavefronts in the ion foreshock region. The incoherent component is weak in the quasi-peperpendicular magnetosheath and larger in the quasi-parallel magnetosheath, consistent with the expected laminar and turbulent nature of these two magnetosheath regions.  In Fig.~\ref{fig:globalinc}(b), the rotational discontinuity has penetrated the magnetosheath and is moving down the magnetotail. At this time, the previously quasi-perpendicular region has changed to the quasi-parallel geometry, and as a result it has started to develop turbulence. Note also the growth of the coherent components in the new quasi-parallel region.
The movie of this run shows that the head of the turbulence/dissipation region in the previously quasi-perpendicular region, as measured by the incoherent component, follows the front of the rotational discontinuity. Figures~\ref{fig:globalinc}(c,f) show that the extent of the incoherent components as well as the coherent components have gone down in the previously quasi-parallel magnetosheath whereas they have both become more volume filling in the new quasi-pararallel region. From these observations, we conclude that the wavelet technique describes well the dynamical change of the magnetosheath regions from turbulent to laminar and vice versa.  

\begin{figure}
\includegraphics[width = 0.6\textwidth]{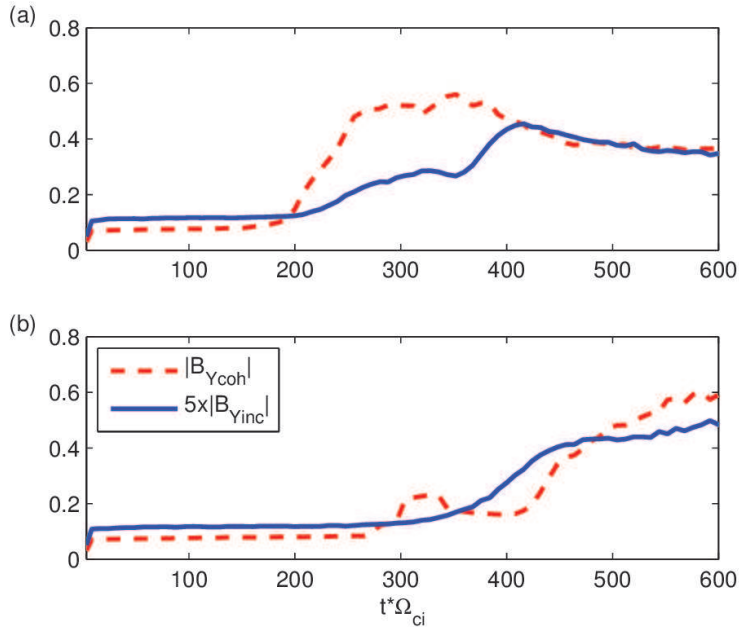}
\caption{The norm of $B_{coh}$ and $B_{inc}$ from the two square sub-domains pictured in Fig.~\ref{fig:globalinc} over the course of the global hybrid simulation. Panel (a) comes from the top box, and (b) the bottom box. \label{fig:boxnorms}}
\end{figure}

To illustrate the change of the turbulent regions over time, we analyze the flows within two small sub-domains of the simulation. These two sub-domains are the 256-cell wide squares plotted in each panel of Fig.~\ref{fig:globalinc}. The magnetic field $B_z$ within each square is de-composed into coherent and incoherent parts over the course of the simulation, and the norms of the signals are plotted in Fig.~\ref{fig:boxnorms}. Box (a) is initially downstream of the quasi-parallel bow shock. Here, the development of turbulence in similar to the Kelvin--Helmholtz unstable flows analyzed previously: a relatively strong coherent signal develops early ({\it e.g.}, at $t=250/\Omega_{ci}$), and the incoherent or turbulent feature grows shortly thereafter. After the rotational discontinuity in the solar wind crosses the planet, Box (a) is then downstream of the less turbulent quasi-perpendicular bow shock. The total level of magnetic fluctuation energy decreases somewhat at this stage (after $t\sim400/\Omega_{ci}$).

For box (b), the transition to turbulence occurs after the rotational discontinuity crosses the planet, and the bow shock becomes quasi-parallel on the southern hemisphere. Because of the background solar wind flow, turbulent fluctuations are advected into box (b) even before large coherent waves develop ({\it e.g.}, at $t=400/\Omega_{ci}$). This highlights that in systems with non-local drives and strong convective contributions such as the foreshock (which derives its free energy from particles reflected from a faraway shock and is embedded in a high-speed solar wind flow), the transition to turbulence need not follow the precise pattern found in the somewhat simpler sheared flow layers. The sheared flow layers exhibited a classical cascade from the large scales of the global flow to smaller scales as turbulence developed locally. In observational data, however, turbulent fluctuations at a given spatial location may precede in time the observation of large coherent structures.

\section{Summary}
\label{sec:summary}

An iterative wavelet filtering technique was applied to a set of simulations of Kelvin--Helmholtz unstable plasma flows to separate the current density into coherent and incoherent pieces. As the global scale Kelvin-Helmholtz vortices developed, a large coherent signal was extracted representing the current sheets on the boundaries of each vortex. As secondary tearing and other processes induced a cascade to smaller scales, the flow transitioned to a turbulent state. The onset of turbulence over time was identified by a sharp increase in magnitude of the incoherent background. While there is no generally accepted definition of turbulence, it is commonly thought that turbulence is not deterministic and is associated with a degree of randomness. This view is supported by our demonstration that the development of turbulence is associated with a sharp increase in magnitude of the incoherent background, which has a noise-like structure.  This diagnostic proved effective for a large fully kinetic simulation as well as a set of three smaller simulations that employed different physical models (fully kinetic, hybrid kinetic ion/fluid electron, and Hall MHD). 

An interesting correlation was found between the increase of the incoherent background and the increase of the plasma thermal energy, both of which display a sharp uptick as strong turbulence develops. Furthermore, while energy conversion in kinetic turbulence appears to be localized to regions of coherent intermittent structures~\cite{wan:2012,karimabadi:2013,karimabadi:2014}, these regions tend to be co-located with the strongest incoherent background. Together these facts suggest that the incoherent background could play a role in the dissipation of turbulent kinetic energy, a possibility we plan to explore in future work. 

In addition, we investigated the application of the wavelet-based diagnostic to a more complex system: a global hybrid (kinetic ion/fluid electron) magnetopsheric model. The model included a rotational discontinuity launched in the incoming solar wind, which changes the orientation of the magnetic field and moves the location of the more turbulent quasi-parallel bow shock region. The wavelet technique efficiently diagnosed the upstream coherent foreshock waves as well as the downstream turbulent regions characterized by incoherent fluctuations spread across a range of scales. The localization of the wavelet modes (as opposed to Fourier modes that are spread evenly across space) was essential for characterizing the turbulence in this inhomogeneous system. An important complication is also illustrated by the analysis of the bottom box in the global magnetosphere model: incoherent fluctuations may exist even before large coherent structures develop \cite{cerri:2017}. Note that pre-existing noise may feed back onto the evolution of coherent structures, such as the tearing of developing current sheets into plasmoids or magnetic islands \cite{comisso:2017}.

The ratio of incoherent to coherent component $R_{ic}$ in our examples varies in a relatively small range of 0.07 to 0.25. The latter value, which occurs in our PIC simulations, is most likely too high for most applications and is affected by particle noise. Thus the real range for $R_{ic}$ is expected to lie in a tighter range.  As such, $R_{ic}$ provides an approximate threshold for turbulence onset. It is a remarkable fact that in fluid turbulence, a single parameter $R_e$ (Reynolds number), which is based on the system properties, is a predictor of whether the system will develop turbulence. There is as yet to be found such a parameter in plasmas. It is tempting to draw an analogy between $R_{ic}$ and the fluid Reynolds number $R_e$. However, there is a big distinction. Unlike $R_e$, our parameter $R_{ic}$ cannot determine a priori based on system parameters whether the system would remain laminar or develop turbulence. Rather $R_{ic}$ indicates that if the system could reach $R_{ic} \sim 0.1$, then it would transition to turbulence.


\begin{acknowledgments}
A.L. was supported by the LDRD office at LANL and thanks Bill Daughton for helpful discussions. 
K.S. acknowledges support by the French Research Federation for Fusion Studies
carried out within the framework of the European Fusion
Development Agreement (EFDA).
V.R.'s contributions were supported by NASA grant NNX15AR16G.
A.S. and L.C. acknowledge support from the Applied Mathematics Research Program of the Applied Scientific Computing Research Office in the U.S. Department of Energy Office of Science.
Simulations were performed on Pleiades provided by NASA's HEC Program and LANL Institutional Computing resources. This work utilized additional resources provided by Blue Waters sustained-petascale computing project, which is supported by the National Science Foundation (awards OCI-0725070, ACI-1238993) and the state of Illinois. Blue Waters allocation was provided by NSF through PRAC award OAC 1614664.
\end{acknowledgments}

\appendix*
\section{Noise in Particle Simulations}

For the fully kinetic PIC simulation and the hybrid PIC simulation in Figs.~\ref{fig:jnorms}(a-b), the incoherent piece of the current density has non-negligible norm even at the beginning of the simulations. This offset of the incoherent piece is produced not by some initially imposed level of turbulence, but rather by the presence of numerical noise associated with the PIC method. In PIC kinetic modeling, which samples phase space with a finite number of numerical macro-particles, there is statistical numerical noise of the current density $\propto 1/\sqrt{N_p}$, where $N_p$ is the number of macro-particles in each grid cell of the domain. The particle noise results from statistical fluctuations in the number of particles in each cell, and it therefore has spatial features on the scale of the grid. Below, we explore the use of wavelet filtering for de-noising particle simulations and compare it to other smoothing algorithms. A detailed study of wavelet-based de-noising for density estimation can be found in \cite{nguyen:2010}.

\begin{figure}
\includegraphics[width = 0.8\textwidth]{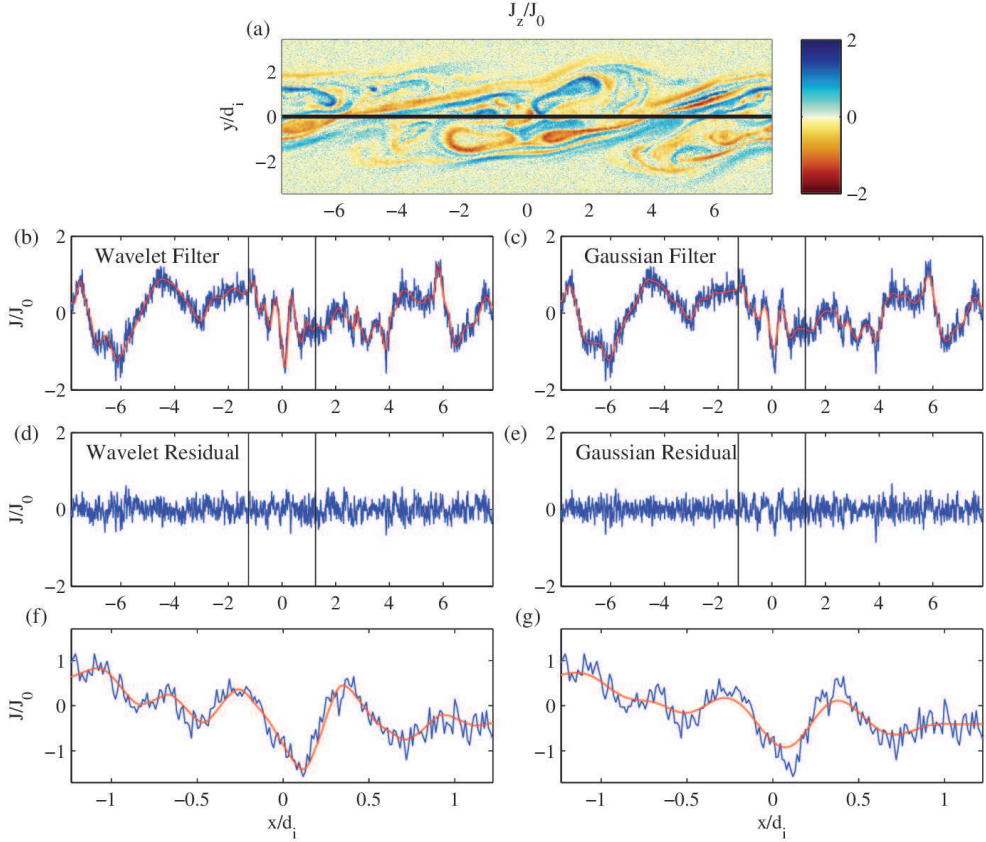}
\caption{(a) The out-of-plane current density $J_z$ from the VPIC fully kinetic simulation with 150 particles per cell per species. The current density along the horizontal cut in (a) is plotted in blue along with filtered data in red based on (b) an iterative wavelet filter and (c) a Gaussian filter. (d-e) Show the residuals (current density minus filtered data). (f-g) Plots zooming in on the region delineated by vertical lines in (b-e). Note that the wavelet filter tends to maintain peak values in current sheets on electron scales, while a Gaussian filter tends to smooth these out. \label{fig:cuts}}
\end{figure}

The out-of-plane current density $J_z$ plotted in $Fig.~\ref{fig:cuts}(a)$ is computed from summing the contributions from particles in each cell of the simulation, which in this case was initialized with 150 particles per cell. The grid-scale numerical noise is apparent. In Fig.~\ref{fig:cuts}, two methods for filtering the grid-scale statistical noise for a PIC simulation are compared. The first is the wavelet filter applied above to study coherent turbulent structures. The only difference is that a value of the multiplicative factor $\alpha=1$ is used. The second filtering method is a classical low-pass Gaussian filter, which convolves the signal with a Gaussian kernel.  A cut of $J_z$ along the center at $y=0$ is plotted (in blue) in Figs.~\ref{fig:cuts} (b) and (c) along with 1D filtered data (in red). The wavelet filtered data $\tilde{J_z}$ was obtained using the iterative method described above. The residual noise, $J_z - \tilde{J_z}$, is plotted in Fig.~\ref{fig:cuts} (d). The width of the Gaussian filter, $\sigma\sim6$ cells, in Fig.~\ref{fig:cuts}(c) was chosen so that the noise extracted (the residuals) in Figs.~\ref{fig:cuts}(d) and (e) have the same norm.

The wavelet filter and the Gaussian filter result in similar de-noised signals. The largest differences between the two filtering methods occur at narrow current sheets. Figures~\ref{fig:cuts}(f) and (g) zoom in on the regions between the vertical lines in Figs.~\ref{fig:cuts}(b-e). A main advantage of the wavelet filtering method is that it better preserves the peak values of sharp features in the current profile. By design, the wavelet basis captures significant features at any scale. The Gaussian filter (or similarly any low-pass band filter), on the other hand, preferentially smooths out small-scale features. The peak values of the current density in the thin sheets in this region are therefore substantially reduced by the Gaussian filter.

\begin{figure}
\includegraphics[width = 0.5\textwidth]{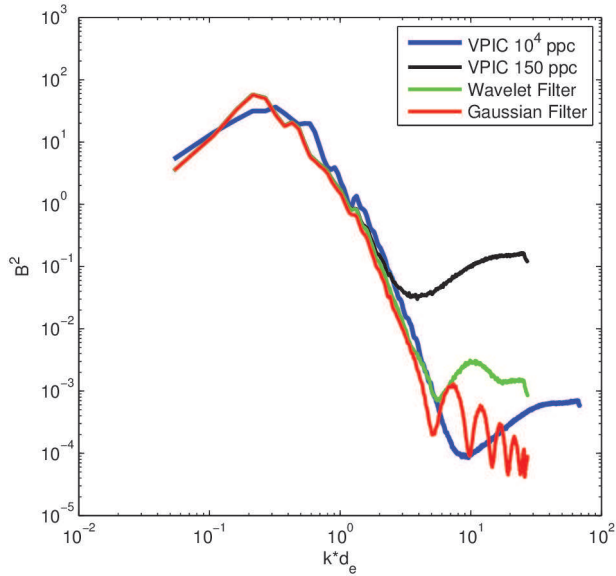}
\caption{A wavelet filter and a low-pass Gaussian filter were used to remove particle noise from a VPIC simulation with 150 particles per cell. The resulting magnetic field spectra are compared to a high-resolution VPIC run with 10,000 particles per cell.\label{fig:wspect}}
\end{figure}

The wavelet and Gaussian filtering provide means of de-noising by post-processing the PIC data after a run. We compare the effect of de-noising through post-processing to runtime methods that are intrinsically less noisy. One method of reducing noise is simply to increase the number of particles in the simulation, which results in smaller statistical noise $\propto1/\sqrt{N_p}$ but increased computational cost. In Fig.~\ref{fig:wspect}, we include a spectrum of magnetic fluctuations from a higher-resolution VPIC simulation with 10,000 particles per cell. The spectrum may be compared to the lower-resolution VPIC run with 150 particles per cell, as well as data from the lower-resolution run de-noised with either a Gaussian filter or the iterative wavelet technique. For the unfiltered data, the spectra turn upwards at large $kd_e> 5$, which corresponds to roughly the grid scale. The higher-resolution simulation with 10,000 particles per cell has reduced noise, and the portion of the spectrum unaffected by particle noise extends to higher $k$ than in the case with only 150 particles per cell. 

\end{document}